\documentclass[]{aastex7}

\usepackage[]{natbib}
\usepackage{amsmath}
\usepackage{caption}
\usepackage{hyperref}
\usepackage{graphicx}

\begin{document}

\newcommand{\vdag}{(v)^\dagger}
\newcommand\aastex{AAS\TeX}
\newcommand\latex{La\TeX}
\newcommand{\jss}[1]{{\bf\textcolor{purple}{#1}}}

\shorttitle{\texttt{FluxCT}}
\shortauthors{Dong et al.}


\title{\texttt{FluxCT V2.0}: Updates to a Web Tool Identifying Contaminating Flux in Space Telescope Data}

\correspondingauthor{Zilin Dong}
\email{zilin.dong@vanderbilt.edu}

\author[0000-0000-0000-0000]{Zilin Dong}
\affiliation{Department of Physics and Astronomy, Vanderbilt University, Nashville, TN 37235, USA}
\email{zilin.dong@vanderbilt.edu}

\author[0000-0002-1043-8853]{Jessica Schonhut-Stasik}
\affiliation{Department of Physics and Astronomy, Vanderbilt University, Nashville, TN 37235, USA}
\affiliation{Neurodiversity Inspired Science and Engineering Fellow, Frist Center for Autism and Innovation}
\email{jessica.s.stasik@vanderbilt.edu}

\author[0000-0002-3481-9052]{Keivan Stassun}
\affiliation{Department of Physics and Astronomy, Vanderbilt University, Nashville, TN 37235, USA}
\email{keivan.stassun@vanderbilt.edu}

\begin{abstract} 
We present \texttt{FluxCT} V2.0, an updated web tool for identifying contaminating flux in Kepler and TESS target pixel files. \texttt{FluxCT} V2.0 focuses on enhancing functionality, user experience, and data processing capabilities. We resolved existing issues to allow for an extended user base, removed known bugs, and extended the tool to any TESS pixel file, allowing the user to search any TESS point object. A batch code for TESS is now available on the companion GitHub. Additional output parameters, such as amplitude dilution and a magnitude cut, have been added to the tool, allowing users more freedom to analyze the possible effects of target sources.
\end{abstract}


\section{Introduction} \label{sec:intro} 
The first edition of \texttt{FluxCT} \citep{FluxCT} was released as a web tool and batch code for the Kepler \citep{Borucki2010} data in January 2023. The popularity of the web tool has led to necessary updates to allow for broader availability and a smoother front-end experience. User feedback has resulted in an additional extension for the TESS \citep{Ricker2015} data, detailed in this paper, the addition of the TESS batch script, and new output parameters. 

The motivation for \texttt{FluxCT} is to allow users to investigate contaminating flux introduced by undetected sources in data from space telescopes with large pixels, i.e., Kepler with 4\farcs{0} pixels and TESS with 21\farcs{}. The integration of flux over multiple pixels, known as Target Pixel Apertures (TPA), produces the light curves from which outputs such as planetary transits, asteroseismic oscillations, and flares are investigated. Contaminating sources can introduce spurious flux into these lightcurves, causing issues with the accuracy of derived parameters such as planet radii \citep{Atkinson2017}, the misattribution of flares \citep{Jackman2021}, and potentially the observation of asteroseismic oscillations \cite{SchonhutStasik2017, SchonhutStasik2020}. One can also use this tool for the study of binarity. 



\begin{figure}
\includegraphics[scale=0.4]{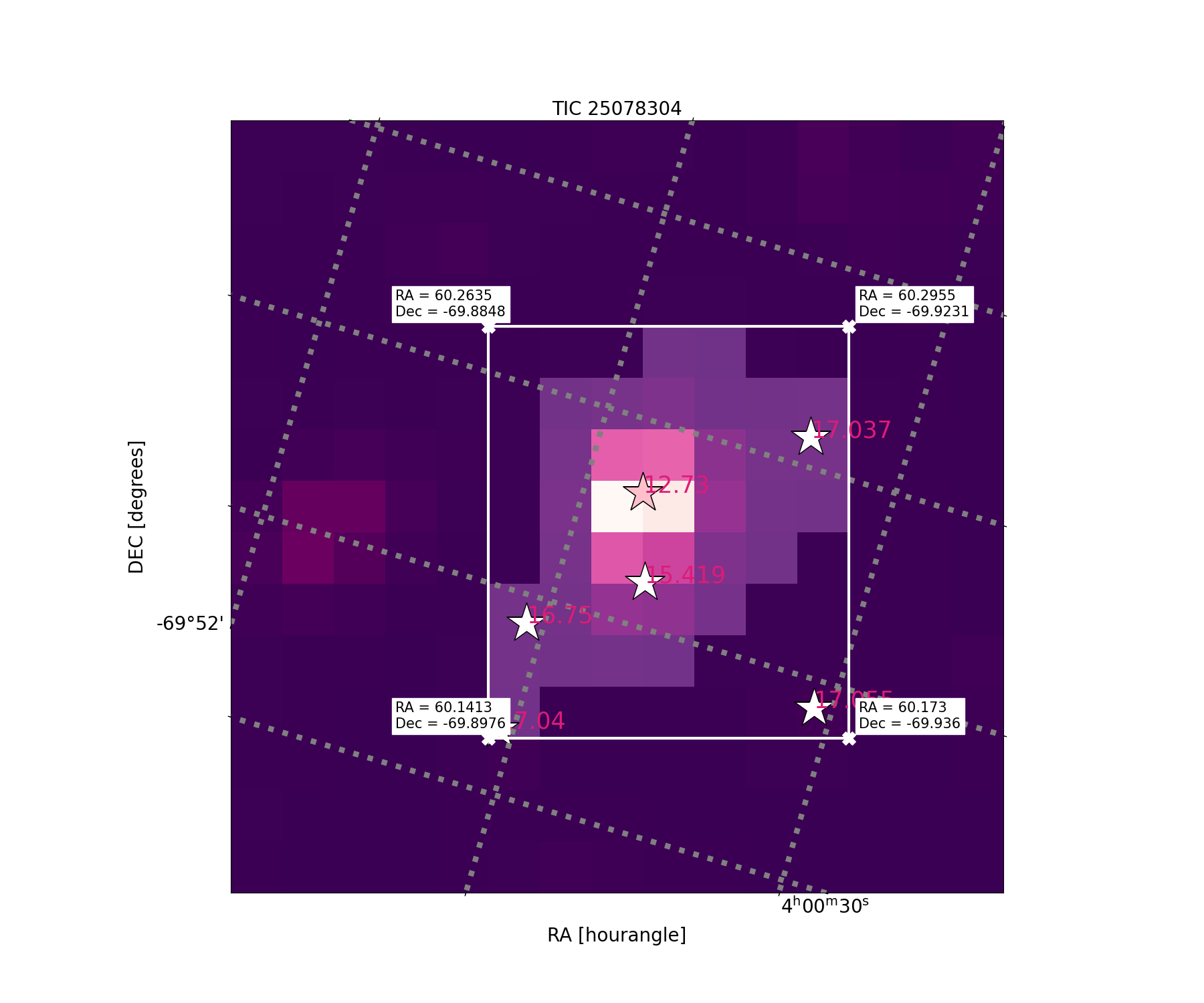}
\caption{Returned plot for TIC 25078304, in the coordinate space RA [hour angle] and Dec [degrees]. The pink star shows the target source, and the white stars correspond to those found in the Gaia search. Grey dotted lines represent on-sky coordinates; each colored square depicts a pixel. Dark purple represents pixels not included in the original TPA, and lighter shades through white correspond to the source's TPA, scaled toward white as flux values increase. White lines around the TPA correspond to the smallest possible polygon, with RA and Dec values in each corner. TIC 25078304 (magnitude 12.73) has five additional sources within the TPA, with target magnitudes of 15.419, 17.037, 17.04, and 17.055 for the accompanying sources, giving magnitude differences between $\sim$3 and $\sim$5. In this case a magnitude cut has been applied so that any stars with a magnitude difference between the source and the target $>$5 are removed from the analysis. The total flux contamination for this system is 16.2\%.}
\label{fig:kic_plot2}
\end{figure}

\section{Web Tool Updates and New Capabilities} \label{sec:webtool} 

To update \texttt{FluxCT}, we combined new \texttt{Python} scripts into the existing Python Anywhere\footnote{\url{https://www.pythonanywhere.com}} framework, allowing users browser access to retrieve data for individual stars. Using the \texttt{FluxCT V2.0} web tool requires only an internet connection and browser access. To run multiple stars, the \texttt{FluxCT} Github contains the relevant script known as the batch code.  

As in the previous version \texttt{FluxCT  V2.0} attempts to pull the source's Target Pixel File (TPF) and Target Pixel Aperture (TPA) using \texttt{lightkurve} \citep{Lightkurve2018}. For Kepler targets, \texttt{FluxCT  V2.0} will use the TPA and create a unique search polygon containing the TPA with as few adjacent pixels as possible. However, TESS data does not always support this method, so we adjusted the \texttt{FluxCT V2.0}'s TPF-generating pipeline. For TESS targets, if a target search does not recover the corresponding TPA by searching the SPOC processing pipelines, \texttt{FluxCT V2.0} will switch to the TASOC pipeline and combine \texttt{TessCut} to create the TPF. 

After the calculation of the polygon, no further deviations occur from the original code. \texttt{FluxCT  V2.0} searches this area using \texttt{astroquery} \citep{Ginsburg2019} (which utilizes \texttt{astropy} \citep{Astropy2022}) to mine the \textit{Gaia} DR3 database \citep{Gaia2021} for any potentially contaminating sources.

A noteworthy enhancement for \texttt{FluxCT V2.0} is expanded support for TESS. Initially, \texttt{FluxCT} was limited to supporting approximately 10\% of TIC IDs. \texttt{FluxCT} V2.0 is capable of processing any TESS point object reduced by the TASOC and SPOC pipelines, markedly increasing its utility by allowing users access to 164.7 million more stars. Additionally, \texttt{FluxCT V2.0} now also outputs the size of Target Pixel Aperture (TPA) in pixels, which over 100 stars, is calculated to be approximately 5 x 5 pixels (corresponding to 105\farcs{} along each axis). For K2 stars, the average TPA size when averaging over 100 EPIC sources was 15 pixels by 13 pixels (corresponding to 60\farcs{} by 52\farcs{}. Additionally, TIC stars within the TPA with a magnitude difference $\ge5$ from the target star will not be returned. 

Another significant addition to the tool's capabilities is the calculation of amplitude dilution. This feature addresses a critical aspect of analyzing light curves—quantifying the extent to which the amplitude of a star's observed flux becomes reduced due to the diluting effect of additional light in the TPA. This enhancement not only improves the accuracy of the data analysis but also broadens the scope of research applications for \texttt{FluxCT}, allowing for more precise interpretations of stellar variability and exoplanet transits. Using a test sample of 200 TESS stars and 100 K2 stars, we found an average amplitude dilution of 39.258\% and 0.404\% correspondingly. The significant disparity between K2 and TESS is consistent with expectations, as the TESS pixels are five times larger than Kepler.

\texttt{FluxCT V2.0} continues to provide all the capabilities of the first version, which are detailed in CITE. 

\section{Version 2.0 Benefits} \label{sec:use_case} 


Expanding the tool's capabilities to include all TIC stars significantly broadens the possible scope of research and analysis. Adding features such as the calculation of the TPA size and amplitude dilution provides deeper insight into the potential contaminating effects and the extent of flux dilution in the light curves, leading to more accurate and reliable data interpretations. 

With the introduction of \texttt{FluxCT V2.0}, the tool's interface remains unchanged and continues to be user-friendly. However, users may need to familiarize themselves with some changes in the tool's operation, especially those who have previously used the older version. Users should ensure they are utilizing the updated batch code to take full advantage of this feature. Additionally, the time required to produce each TIC star using TESSCut and TASOC/SPOC pipelines might make running the batch code slightly lengthy (200 TIC took 220 minutes).

\section{Accessibility} \label{sec:access} 
\texttt{FluxCT V2.0} is currently available as a web tool that allows the search of single targets (\url{http://jstasik.pythonanywhere.com}). Creating an accessible web tool is motivated by the desire to encourage broader access to space telescope data for students. The ease of use allows students as early as high school to explore the data without needing advanced knowledge of \texttt{Python}. A full version that can search multiple sources is available at the GitHub \url{https://github.com/ZilD117/FluxCT-V2.0}. The corresponding author welcomes any suggestions for updates. More example plots and supplemental material exist at \url{https://www.jessicastasik.com/flux-contamination-tool}. A frozen version of the code is available on Zenodo \cite{Dong2024}. 

\begin{acknowledgments}
This paper includes Kepler mission data obtained from the MAST data archive at the Space Telescope Science Institute (STScI). 
We make use of data from the European Space Agency (ESA) mission \textit{Gaia} (\url{https://www.cosmos.esa.int/gaia}), processed by the \textit{Gaia} Data Processing and Analysis Consortium (DPAC,
\url{https://www.cosmos.esa.int/web/gaia/dpac/consortium}).

\end{acknowledgments}

\bibliographystyle{aasjournal}
\bibliography{sample631}

\end{document}